Klaus Gottstein

# New insights?
# Heisenberg's visit to Niels Bohr in 1941 and the Bohr letters[1]

The documents recently released by the Niels Bohr Archive do not, in an unambiguous way, solve the enigma of what happened during the critical brief discussion between Bohr and Heisenberg in 1941 which so upset Bohr and made Heisenberg so desperate. But they are interesting, they show what Bohr remembered 15 years later. What Heisenberg remembered was already described by him in his memoirs "Der Teil und das Ganze". The two descriptions are complementary, they are not incompatible. The two famous physicists, as Hans Bethe called it recently, just talked past each other, starting from different assumptions. They did not finish their conversation. Bohr broke it off before Heisenberg had a chance to complete his intended mission.

Heisenberg and Bohr had not seen each other since the beginning of the war in 1939. In the meantime, Heisenberg and some other German physicists had been drafted by Army Ordnance to explore the feasibility of a nuclear bomb which, after the discovery of fission and of the chain reaction, could not be ruled out. How real was this theoretical possibility? By 1941 Heisenberg, after two years of intense theoretical and experimental investigations by the drafted group known as the "Uranium Club", had reached the conclusion that the construction of a nuclear bomb would be feasible in principle, but technically and economically very difficult. He knew in principle how it could be done, by Uranium isotope separation or by Plutonium production in reactors, but both ways would take many years and would be beyond the means of Germany in time of war, and probably also beyond the means of Germany's adversaries. (When Heisenberg heard about the Hiroshima bomb, almost four years later while interned in Farm Hall, at first he could not believe it.) Heisenberg and von Weizsäcker were very glad about this result. It meant that they were spared a difficult moral decision. They were able to concentrate on the construction of a reactor for power production, a goal easily compatible with their value system. If a bomb would have been within reach, how could they have avoided being forced to build it without sacrificing their lives as saboteurs? But what about the long-range future? Was the construction of nuclear weapons unavoidable? Was it conceivable that the then small community of nuclear physicists could come to an agreement not to work on bomb construction? Heisenberg and his friend and colleague von Weizsäcker decided that it would be helpful to have discussions with Bohr, their father figure. In a kind of naiveté they did not realize that their old cordial relationship with Bohr had been affected by the events of the war. For Bohr his old friend Heisenberg was now a representative of an enemy country, of the occupying power of his native Denmark, whose remarks would have to be looked upon with suspicion.

---

[1] Klaus Gottstein was a member of the Max Planck Institute for Physics from 1950 to 1970 under the directorship of Werner Heisenberg. For several years he was head of the experimental division of the Institute. In 1969 he asked Heisenberg about his visit to Bohr in 1941, and Heisenberg told him. Immediately afterwards K.G. dictated into a recorder what Heisenberg had said. This record and its transcription are still available.

Heisenberg managed to make the trip to Copenhagen in September 1941, using the opportunity of a scientific conference arranged by the German Culture Institute in Copenhagen, boycotted by Bohr. Heisenberg spent several days in Copenhagen and probably saw Bohr several times, in Bohr's office, in Bohr's home and on a walk. On the latter occasion when there was no danger of being overheard by the Gestapo, Heisenberg undertook to broach the questions which were the real reasons for his trip.

What is written above is the gist of what Heisenberg wrote and what he explained to friends and colleagues when questioned. But his explanations were not accepted everywhere, particularly not by some of his British and U.S. colleagues and by some later writers who were convinced that Heisenberg did all he could to make the bomb for Hitler, but failed, and after the war tried to white-wash himself. The situation became even worse in 1956 when the American journalist Robert Jungk published a book "Heller als tausend Sonnen" (Brighter Than a Thousand Suns) in which he described, greatly exaggerating, Heisenberg's satisfaction with the technical difficulties of bomb construction and the lack of enthusiasm for overcoming these difficulties, as a secret plan to prevent, for moral reasons, the construction of an atomic bomb for Hitler which otherwise he could have built. Heisenberg, and particularly von Weizsäcker, wrote letters to Robert Jungk in which, while appreciating Jungk's extensive research and detailed accounts of the developments, criticized some of his generalisations and exaggerations. Cathryn Carson, in her article "Reflexionen zu 'Kopenhagen'", appended to the German edition of Frayn's play "Copenhagen", quotes from these letters. In the Danish translation of his book, which appeared in 1957, Jungk published an extraction of Heisenberg's letter, but only the laudatory part. It was known that Bohr took exception to Jungk's book which he had read in the Danish edition. Jungk's book, unfortunately, did much to harm Heisenberg's credibility, particularly as the wrong impression had arisen in some quarters that Heisenberg had commissioned it. Heisenberg was unaware of this. He never "portrayed himself after World War II as a kind of scientific resistance hero who sabotaged Hitler's efforts to build a nuclear weapon", as was suggested by James Glanz in The New York Times recently. Heisenberg always stressed how content he had been that nuclear weapons did not seem to be feasible for several years to come so that Hitler and his government made no efforts to build them when this had become clear to them.

Meanwhile, all kinds of rumours circulated about the "real" motives behind Heisenberg's 1941 visit to Bohr. It was suggested that he wanted to do some spying, to find out what Bohr knew about the nuclear efforts in the U.S. and Great Britain. It was suspected that Heisenberg wanted to enlist the support of Bohr for the German project. On the basis of some conversations which Heisenberg and von Weizsäcker had had with members of Bohr's institute there was also the version that the real reason for Heisenberg's visit was the intention to convince Bohr that Germany was going to win the war, that this outcome was desirable, and that Bohr had better end his unwillingness to cooperate with German authorities. It was generally held that the formerly cordial relationship between Bohr and Heisenberg was severely disturbed, if not severed, ever since. Few members of the international community knew that they continued to have friendly relations after the war, visiting each other, with their families, in their homes in Copenhagen and Göttingen, spending their vacations together in Greece, and that Bohr wrote an article for the Festschrift to Heisenberg's sixtieth birthday in 1961.

When it became known that the Niels Bohr Archive in Copenhagen held a letter by Bohr to Heisenberg, written after the appearance of Jungk's book but never sent, speculation concentrated on this document, to be published 50 years after Bohr's death, i.e. in 2012, from which the solution of all the open questions was expected. However, to end speculation, the Niels Bohr Archive released 11 documents pertaining to Heisenberg's visit, including the much-discussed unsent letter, preceded by an article by Aage Bohr, published in 1967, on "The War Years and the Prospects Raised by Atomic Weapons". The documents, with the exception of one letter written by Heisenberg to Bohr, are unfinished drafts written by Bohr in the late 1950s and early 1960s, addressed to Heisenberg, but never sent. As the director of the Niels Bohr Archive, Finn Aaserud, points out, the documents have to be viewed with caution. They were written 16 years or more after the event and represent just drafts, not finished papers. Nevertheless, the contents of the documents are interesting and, depending on the pre-established views and opinions of the readers of today, surprising to a lesser or greater degree. Here are some of the general characteristics of the documents, with my comments in brackets:

- Bohr's tone in addressing Heisenberg is extremely cordial and friendly.
- Bohr was still highly interested in clarifying Heisenberg's intentions and motivations behind his 1941 visit. His sentences in Document 11 c "I have long been meaning to write to you ..." and "I have written in such length to make the case as clear as I can for you and hope we can talk in greater detail about this when opportunity arises" are proof of this. (This is new information. Heisenberg was under the impression that Bohr and he, having differing recollections of their discussion, had come to the conclusion that it would be best to let rest the spirits of the past. It is a pity that the letter was not sent. Several opportunities for clarifying conversations were missed at later meetings of Bohr and Heisenberg. It seems that Bohr was afraid he might hurt Heisenberg's feelings by insisting too much on his interpretation of the events.)
- Document 1 contains the confirmation that Bohr and Heisenberg met several times during Heisenberg's visit to Copenhagen in 1941: Bohr refers to "our conversations" in the plural, and he mentions "our conversation in my room at the institute" as well as the strong impression Heisenberg's remarks made "on Margrethe and me". Since it is unlikely that Bohr's wife Margrethe was present at the confidential conversation in Bohr's room in the institute one may assume that Heisenberg's recollection is correct that he was also invited to Bohr's home. Moreover, there is Heisenberg's and von Weizsäcker's testimony that the critical discussion took place during a walk, to avoid unwanted earwitnesses.
- Bohr was, at the time of the visit in 1941, highly distressed by the circumstances of Heisenberg's visit, his lecture at the German Culture Institute and his contacts with the German Embassy (more correct: Legation) in Copenhagen. (Heisenberg had assumed Bohr would understand that without such contacts he would not have obtained visa and permission to enter occupied Denmark.)
- Bohr understood and appreciated that one of Heisenberg's reasons for the visit was genuine care: to see how Bohr and his institute fared under German occupation and to be of assistance, if at all possible (Document 11 c).
- For Bohr it was of central and sad significance that Heisenberg during his visit expressed his conviction of a German victory whereas Bohr, as a Danish patriot, had placed all his hopes in a German defeat. Since towards the end of the war Heisenberg's conviction must have disappeared, Bohr wondered whether Heisenberg, in retrospect, had forgotten or repressed his earlier views. (Again, for Heisenberg, mentioning the prospects for a German victory, was not central to his mission. At the beginning of the war he had, in private, expressed the view that Hitler would lose the war like a chess-player would lose a game into which he entered with one castle less than his opponent. However, after the surprisingly fast defeats of Poland and France, the occupation of large parts of Europe and the initial great victories and advances in the Soviet Union, with the U.S. still neutral, Heisenberg like most non-nazi Germans had come to the conclusion that a German victory now seemed likely. They feared that a German defeat would mean Soviet occupation of Europe which, even for anti-nazis, was considered an even greater evil than German domination. Auschwitz and the full extent of nazi crimes was not yet known, but Stalin's massacres were. The hope - completely unrealistic as we now know but considered realistic at the time - was that after a German victory the German army would get rid of Hitler and his henchmen. The anti-nazi stance of many German generals, who later took part in the assassination plot of July 20, 1944, was known to persons who, like Heisenberg through the "Wednesday Society", were close to opposition circles. For Heisenberg, it was part of his care for Bohr to think in sober terms of the future and of Bohr's and his institute's survival. It would be advisable to end opposition to a victorious Germany. It is obvious that Heisenberg's assessment of Germany's chance to win the war must have changed a few months later when the U.S. entered the war and the German army suffered severe setbacks in Russia.)

- Bohr mentions several times his reticence caused by his suspected surveillance by German police. (There is no indication of an awareness by Bohr that Heisenberg was under the same handicap. He had to be extremely cautious in choosing his language. Mentioning to Bohr the existence of a German nuclear programme and of his involvement in it, could be interpreted, and probably was, treason punishable by death. In public conversations, also in the cafeteria of Bohr's institute, he may have had to say things which did not represent his opinion. This situation is well-known to people having lived under cruel dictatorships.)

- Document 6 says that Heisenberg "did not wish to enter into technical details but that Bohr should understand that he knew what he was talking about as he had spent 2 years working exclusively on this question." Bohr had known about the possibility of nuclear weapons only in a very general way and at that time still had held the opinion that the technical difficulties were insurmountable. Therefore Heisenberg found it necessary to mention his two years of investigations in order to convince Bohr that he was not "talking moonshine". Bohr had been "doubtful looking" (Document 11 a).

- Bohr wondered (and this is new information) who had authorized Heisenberg to discuss with him military secrets. (Heisenberg had assumed Bohr would understand that he spoke in his private capacity as Bohr's old friend and colleague who, however, because of the delicacy of the subject discussed, had to use very involved language. Bohr, on the other hand, could not imagine that Heisenberg acted on his own initiative, without any special permission, let alone orders. But this was so. Heisenberg had thought, naively, that Bohr would be ready, as he always had been in earlier times, to discuss with him possible solutions for complicated problems. He had lacked the sensitivity for Bohr's patriotic feelings and misgivings under the changed circumstances of war and occupation. On the other hand, it is justified to say that it took great moral courage to talk to Bohr about implications of his secret work. Heisenberg risked his neck.)

- Constant German propaganda talks of the imminent use of "new weapons" fortified suspicions by Bohr and his Danish colleagues that there was a German nuclear bomb programme. Assertions by Jensen to the contrary were not trusted though he himself was considered honest. But Jensen was working on the reactor programme, and it had to be doubted that he was privy to all aspects of the programme.

- After Bohr's escape to Sweden and subsequent flight to Great Britain in the autumn of 1943 "it was quite clear already then, on the basis of intelligence reports, that there was no possibility of carrying out such a large undertaking in Germany before the end of the war". (Document 11 b). This is a remarkable confirmation of Heisenberg's own conclusion. It is also interesting that these intelligence reports had no influence on the progress of the Manhattan project.

- Aage Bohr writes "After the outbreak of war and especially after the occupation of Denmark we in Copenhagen were completely cut off from following the allied nations' efforts in the field of atomic energy." Niels Bohr confirms this in Document 11 c. (Also Heisenberg knew that. How could he expect to do some spying, as some writers have suggested?) In a footnote to his article Aage Bohr assures the reader that no secret plan was submitted to his father by Heisenberg "aimed at preventing the development of atomic weapons through a mutual agreement with colleagues in the allied countries." Again, in Document 11 c, this is what Bohr remembers. It is quite true, also according to Heisenberg. It had indeed been Heisenberg's intention to get Bohr's opinion on possibilities for such an agreement or on other ways out of the impasse presented by the basic feasibility of atomic weapons. But Heisenberg never had a chance to present his questions because of Bohr's reticence and Bohr's unwillingness to continue the conversation when Heisenberg, as an introduction, had told Bohr that atomic weapons were technically possible, and that he knew it. He was not even allowed to add, as he had intended, that the technology was very difficult and would take a long time, thereby giving the small international community of atomic scientists a chance to use their influence in the meantime. Bohr had stopped listening. This is admitted by Bohr in Document 11 c where he writes "During the conversation, which because of my cautious attitude was only brief ...".

New insights